\documentstyle[aps,pre,twocolumn,epsfig]{revtex}
\title{Broken space-time symmetries and mechanisms of
rectification of ac fields by nonlinear (non)adiabatic
response}

\author{ S. Denisov$^1$, S. Flach$^2$,
A. A. Ovchinnikov$^2$, O. Yevtushenko$^3$ and Y. Zolotaryuk$^{4,5}$}

\address{ $^1$ Department of Chemistry, Tel-Aviv Unversity, 
               Tel-Aviv 69978, Israel\\
          $^2$ Max-Planck-Institut f\"ur Physik komplexer Systeme,
         D--01187, Dresden, Germany\\
          $^3$ Abdus Salam ICTP, 34100 Trieste, Italy\\
          $^4$ Section of Mathematical Physics, IMM, Technical University
               of Denmark, Lyngby 2800, Denmark\\
          $^5$ Bogolyubov Institute for Theoretical Physics, 
               National Academy of Sciences of Ukraine, Kiev 03143, Ukraine\\}

\date{\today}

\begin{document}

\maketitle

\begin{abstract}
We consider low-dimensional dynamical systems exposed to 
a heat bath and to additional ac fields.
The presence of these ac fields
may lead to a breaking of
certain spatial or temporal symmetries which in turn cause
nonzero averages of relevant observables. Nonlinear (non)adiabatic
response is employed to explain the effect. We consider a case
of a particle in a periodic potential as an example and discuss
the relevant symmetry breakings and the mechanisms of rectification
of the current in such a system.
\end{abstract}

\pacs{PACS:
%05.45.-a, 05.60.Cd, 05.45.Ac
}

\section{Introduction}

Much has been written on noise induced transport, where noise may
be coloured, or simply white plus time periodic signal.
Arguments include violation of fluctuation-dissipation theorems,
breaking of reflection symmetries of potentials in space, Maxwell
daemons, mixing of harmonics (of e.g. a periodic drive), etc.
All this applied to both classical and quantum systems, and
extended at least conceptually to stochastic resonance, quantum
stochastic resonance, etc. A recent review on ratchet transport by
Reimann \cite{Reimann} provides with a lot of theoretical and
experimental results, and we refer the interested reader to this
work.

For a set of related problems, such as directed particle current 
\cite{symmetry,kinetic}, directed energy current \cite{fzmf01},
average magnetization \cite{fo01pa}, and
nonlinear Hall effect \cite{aao02}, to name a few, a recently published
symmetry approach was shown to systematically account for  
all relevant symmetries which have to be broken in order to explain 
the observed rectification effect. The purpose of this paper
is to generalize this approach, 
and to apply it to different physically relevant situations like
underdamped, overdamped or zero temperature cases.
We also use the nonlinear response arguments
to make the results of symmetry considerations very transparent.
We will argue for a rather minor role of additional fluctuations
which are 
mainly responsible for the area of phase space explored. Although 
we will discuss mostly the case of
time-periodic external fields, we will also
show how the nonlinear response concept
can be generalized to fields which are quasiperiodic in time.

The paper has the following structure. In the next 
section we discuss the symmetry breaking by considering the adiabatic
limit and making use of some general forms of nonlinear response
functions. Section \ref{dircur} is devoted to the detailed symmetry
analysis of a classical particle moving in a periodic potential
under the influence of an external ac drive. Symmetry breaking leads
to a directed current here. We will discuss the presence
of ballistic channels which provide with a mechanism of 
current rectification.
We will show that these ballistic channels survive in the presence
of dissipation.
Section \ref{NonParDR} addresses the case of a particle with
nonparabolic dispersion. 
The addition of a spatial potential allows us to tune
the system in such a way that ballistic channels disappear.
This will lead to 
an expected dramatic drop of the directed current value.
Conclusions and discussions are given in Section \ref{Conclusions}.

%######################################################################
%######################################################################
\section{ Adiabatic response to slow \\
          ac fields
        }
\label{adiab}
%######################################################################
%######################################################################

Consider a certain system in contact with a heat bath.
The system is characterized by some internal (nonlinear) dynamics,
and we will discuss concrete models below. So far we just need to
know that its state can be characterized by certain variables
which are functions of time. These could be functions of phase space
variables of a classical system, or expectation values of operators
for a quantum system. Consider one of such variables which we
denote as $A(t)$. The coupling
to the heat bath will be characterized by at least two parameters -
the temperature of the bath (we will use the notation of inverse
temperature $\beta$) and some set of relaxation times, in the simplest
case just one relaxation time (we will use the notation of the
inverse relaxation time $\nu$). We also assume that the  
chosen variable $A(t)$ is zero on average.

Let us now apply a static field $E$ which couples to $A$ such that a nonzero
average $\tilde A = \langle A(t) \rangle_t$ is generated. Its dependence on $E$
is assumed to be given by some single 
valued response function $\tilde A = f(E)$
\cite{multi-valued}
(the physical meaning of the response function can be different,
electric polarization or magnetization, for example).
The above mentioned symmetries will be connected with corresponding
symmetries of the function $f(E)$. This function will be in general
nonlinear, yet in some cases it may be expanded in a Taylor
series around $E=0$ and start with a linear term (this term describes
then the linear response). There are two
possibilities: the function $f(E)$ is either antisymmetric $f(E)=-f(-E)$
or it is asymmetric $f(E) \neq -f(-E)$.

Before analyzing the case of adiabatically slow
periodic ac fields, we will more rigorously
introduce the
notion of possible symmetries of an arbitrary periodic function
with zero mean.

%######################################################################
\subsection{Classification of symmetries of a periodic function
with zero mean}
%######################################################################

Consider  a periodic function $g(z+2\pi)=g(z)$
having zero mean. First such functions may be symmetric around certain
$z$ values. Without loss of generality this point may be set to
zero and we find $g(z)=g(-z)$. We will use the abbreviation $g_s$ in such
a case. Second such functions may be antisymmetric $g(z)=-g(-z)$
(abbrevation $g_a$). Note that the points around which a function
is symmetric and antisymmetric will be different if the function
possesses both symmetries.
Finally the function may posses shift symmetry (which is also called
antiperiodicity): $g(z+\pi) = -g(z)$ (abbreviation $g_{sh}$).
A zero mean periodic function
may possess none of the above symemtries, precisely one of them or all
three simultaneously.
In particular $g_s$ functions may be expanded in a pure
cosine Fourier series, $g_a$ functions in a pure
sine Fourier series, and $g_{sh}$ functions show up with zero
even Fourier components in their Fourier series expansion.
As a consequence the simplest function $g(z)=\cos (z+z_0)$ possesses all
three symmetries. The function $g(z)=\cos(z) + \cos (2z+z_0)$
does not posses any of the listed symmetries except
for $z_0=0,\pi$ ($g_s$) and $z_0 = \pm \pi/2$ ($g_a$).
A final example $g(z)=\cos (z) + \cos(3z + z_0)$ always possesses
shift symmetry ($g_{sh}$) and in addition may be symmetric {\it and}
antisymmetric for $z_0=0,\pi$.

Let us finally note that most cases under consideration use several
harmonics contained in the drive. However it may be also important to
use pulse sequences. The symmetry considerations can be straightforwardly
applied also to such cases.

%######################################################################
\subsection{Periodic ac fields}
%######################################################################

Let us now assume that the field $E$ is slowly varying in time.
Slowly means that the characteristic time scales of changes of $E(t)$
are much larger than all other time scales in the system. Then we
are dealing with the adiabatic limit of the response to time-dependent
fields. We may use the response $f$ to a static field and simply
insert the slow time dependence: $f[E(t)]$. Consider a time-periodic
field $E(t)=E(t+T)$ with zero average $\tilde E =0$. In general the
corresponding value $\tilde A = 1/T \int f(E(t)) dt$ 
will be nonzero for asymmetric $f$.
However for antisymmetric single-valued 
$f$ the average $\tilde A$ will be zero provided
the function $E(t)$ is either antisymmetric $E(t)=-E(-t)$ or
has shift symmetry $E(t)=-E(t+T/2)$. 

Consider as an example
%------------------------ 1-1 -----------------------------------------
\begin{equation}
E(t)=E_1 \cos t + E_2\cos (2t+\alpha) + E_3\sin (3t+\alpha ')\;\;.
\label{1-1}
\end{equation}
%----------------------------------------------------------------------
For $E_2=E_3=0$ Eq. (\ref{1-1}) 
has all three 
mentioned symmetries of periodic functions.
For $E_3=0$ the function has no symmetries except for
$\alpha=n\pi$ ($n$ integer) where it is symmetric or
for $\alpha=\pi(1/2+n)$ where the function is antisymmetric.
For $E_2=0$ Eq. (\ref{1-1}) has shift symmetry except for
$\alpha ' = \pi(1/2+n)$ when it is in addition symmetric
and antisymmetric (around different time origins, of course).

Next we assume a case when the adiabatic response function $f$ may
be expanded in a Taylor series:
%----------------------------------------------------------------------
\begin{equation}
f(E)=f_1E + f_2E^2 + f_3E^3 + ...
\label{1-2}
\end{equation}
%----------------------------------------------------------------------
where the skipping of higher order terms is justified by the
smallness of $E$. For $f_2 \neq 0$ the response is asymmetric
and consequently any field from (\ref{1-1}) will in general lead to a
nonzero average of $f$. For $f_2=0$ however all shift symmetric
and antisymmetric $E(t)$ will yield zero averaged $f$
(because it leads also to a shift symmetry or antisymmetry of $f$
in time). As an example of mixing of harmonics 
the leading nonzero contribution
for $E_2 \neq 0$, $E_3=0$ equals
%----------------------------------------------------------------------
\begin{equation}
\tilde f = \frac{1}{T} \int_0^T f[E(t)]dt =
 \frac{3}{4}f_3 E_1^2 E_2 \cos \alpha + ...\;\;.
\label{1-3}
\end{equation}
%----------------------------------------------------------------------
In accordance with the above said $\tilde f$ vanishes
for $\alpha=\pi(1/2+n)$. Note that the antisymmetry property of $f(E)$
is linked to some internal symmetries of the dynamical system under
consideration, as will be shown below.

It is important to note that the written above is valid for single-valued
response functions. If the coupling to a heat bath is too weak, 
response functions may contain hysteretic loops. Such cases need
separate discussion \cite{multi-valued}.

%######################################################################
\subsection{Quasiperiodic ac fields}
%######################################################################

We may even consider quasiperiodic driving here. Suppose that we drive
the system with a field $E(t)=e_1(t) + e_2(t)$ where both components
$e_{1,2}$ are time periodic, but with incommensurable periods. The resulting
field $E(t)$ would be then quasiperiodic. If considering low amplitude fields
and expanding the response function to second or third order, the
full dc average of $A(t)$ will be given by a mere sum of the averages
obtained in the presence of only one of the two periodic fields.
Consider e.g. $E(t)=e_1 \cos \omega_{1} t + e_2 \cos \omega_2 t$
and an asymmetric
response. In lowest order in the field amplitudes we obtain
%----------------------------------------------------------------------
\begin{equation}
\tilde f = f_2\left( \frac{1}{2}e_1^2 + \frac{1}{2}e_2^2 \right)
+ f_4\left( \frac{3}{8}e_1^4 + \frac{3}{8}e_2^4 +
\frac{6}{4}e_1^2e_2^2 \right) + ...
\end{equation}
%----------------------------------------------------------------------
In higher orders in the response interference effects appear, so that
products of the field components enter the result. Yet for the lowest order
case ($f_2$) only the sum appears. So we may assume that further
changing the drive by adding more time periodic field components does not
change the main result of a possibility of a nonzero average. 

If we consider an antisymmetric response function $f$, we need
to choose a more sophisticated quasiperiodic drive. Take e.g.
$E(t)=e_1(t) + e_2(t)$ with $e_i(t)=e_{i1}\cos \omega_it
+ e_{i2} \cos(2\omega_i t+\alpha_i)$.
Assuming again that the field amplitude is small, we obtain in lowest
order
%----------------------------------------------------------------------
\begin{equation}
\tilde f = \frac{3}{4}f_3 \left(e_{11}^2e_{12}\cos \alpha_1
+ e_{21}^2e_{22} \cos \alpha _2 \right)\;\;.
\end{equation}
%----------------------------------------------------------------------
Again we find that the periodic components of the quasiperiodic drive
contribute additively in lowest order. Note that each
contribution can be obtained by considering a reduced drive function
consisting only of this part. The corresponding contribution is
clearly related to the symmetry properties of this reduced
drive.

\subsection{Do we need more?}

Once a given problem is considered in the adiabatic limit like discussed
above, the appearance of nonzero averages due to ac fields can be
obtained, and further changing of parameters away from the
adiabatic limit will change numbers, but not the fact of nonzero
averages. Ratchet transport in its general form falls into
such a class of systems. Indeed a particle moving in a periodic
potential $V(x)=V(x+\lambda)$
under the influence of an external colored noise and
dissipation may be taken into the limit of zero temperature
(no noise), slow periodic driving and zero mass (over damped case):
%----------------------------------------------------------------------
\begin{equation}
\dot{x} + V'(x) + E(t) =0\;\;.
\end{equation}
%----------------------------------------------------------------------
The response of $\langle \dot x \rangle$
to a corresponding static field $E$ is well known and
has been discussed in connection with Josephson current-voltage
characteristics (for symmetric potentials). The response is
clearly nonlinear, and antisymmetric if the potential is symmetric
in space. In case of asymmetric ratchet potentials the response is nonlinear
but asymmetric. In the latter case a simple $E(t) \sim \cos t$
signal with large enough amplitude
will generate a nonzero averaged current. In case of symmetric
potentials a drive which is neither antisymmetric nor shift symmetric
is needed.

Nonadiabatic corrections are of interest and importance. E.g.
upon constantly increasing the frequency of a drive, currents
(or other averages) may change sign, or increase or decrease
by orders of magnitude \cite{bh94},\cite{maf02},\cite{vinokur}. 
Especially interesting
are cases when certain (nonadiabatic) parameter limits give
rise to new symmetries. These symmetries may be traced back away
from the corresponding limits and help understand lots of peculiar
features which are observed in such driven systems.

If the adiabatic response simply vanishes, nonadiabatic response
contributions will be not just mere corrections. In such cases
the symmetry considerations of the underlying equations
of motion presented below appear to be the most
direct way to predict rectification effects. One example of such
a case is given in \cite{fo01pa} where the problem of driven quantum
spins in ac magnetic fields is considered. While the adiabatic
limit provides with zero induced magnetization component in y-direction
there, nonadiabatic response terms make this component nonzero, as
expected from symmetry considerations.

%######################################################################
%######################################################################
\section{Directed currents -
the case of a classical particle in a periodic potential}
\label{dircur}
%######################################################################
%######################################################################

In order to make further progress in the understanding of rectification
it is useful to define a model. Here we will consider the
case of a particle of the unit mass
moving in a space-periodic potential $V(x)=V(x+\lambda)$
under the influence
of friction and external forces:
%----------------------------------------------------------------------
\begin{equation}
m \ddot{x} + \gamma \dot{x} - f(x) - \chi(t) = 0\;\;,
\label{7}
\end{equation}
%----------------------------------------------------------------------
where $f(x) = -V'(x)$ and $\chi(t)$ is zero on average.
The mass $m$ is assumed to be equal to one if not stated otherwise.
If $\chi(t)$ is a Gaussian white noise,
the particle undertakes diffusion with zero net current, in accord
with the fluctuation-dissipation theorem. If however $\chi(t)$ contains
correlations (color), then it is known that a nonzero net (dc) current
is possible. In order to understand this result, it is appropriate
to make the correlations in the noise as transparent as possible.
The easiest way to do so is to choose
%----------------------------------------------------------------------
\begin{equation}
\chi(t) = \xi(t) + E(t)\;\;,\;\;E(t)=E(t+T)\;\;,
\label{7-2}
\end{equation}
%----------------------------------------------------------------------
where $\xi(t)$ is a Gaussian white noise and the time-periodic
external field $E(t)$ has zero mean. A probabilistic description of
the system is then possible using a Fokker-Planck equation (see
Appendix A).

In a next radical step we skip the $\xi(t)$ term. The reason for that
is that we will be left with a deterministic equation, whose
symmetries may be studied. These symmetries involving operations
in time $t$ are assumed to hold even in the presence of a Gaussian
white noise, since such a term does not contain temporal correlations.
Note that the general nonintegrability of the resulting equations may 
provide with irregular (chaotic) trajectories.
Thus we are left with an equation of the form
%----------------------------------------------------------------------
\begin{equation}
m\ddot{x} + \gamma \dot{x} - f(x) - E(t) =0 \;\;.
\label{eqm1}
\end{equation}
%----------------------------------------------------------------------

%######################################################################
\subsection{The relevant symmetries and ways of violations}
%######################################################################
In order to characterize the symmetries of Eq. (\ref{eqm1}) we remind
that the phase space dimension is three.
As we look for nonzero average currents which are characterized
by the velocity $\dot{x}$, we have to check whether there exist
symmetries which allow to generate out of a specific trajectory
of (\ref{eqm1}) another one with reversed velocities. The transformations
of interest have to involve a change of sign of $\dot{x}$ leaving its
absolute value unchanged. Thus
we look for transformations which leave Eq.(\ref{eqm1}) invariant and

\begin{itemize}

   \item either change the sign of $ \, x \rightarrow -x \, $ and
         simultaneously shift time $t \rightarrow t+t_0$ 

   \item invert time $ \, t \rightarrow -t \, $ and shift
         coordinate $x\rightarrow x + x_0$.

\end{itemize}

The following symmetries can be identified \cite{symmetry,kinetic}:
%----------------------------------------------------------------------
\begin{eqnarray}
& \, & \hat{S}_a\;\;: \;\; x \rightarrow -x\;\;, \;\; 
t\rightarrow t+\frac{T}{2}\;\;,
\;\; {\rm if} \; \{f_{a}\;,\; E_{sh} \} \;\; ;
\nonumber
\\
& \, & \hat{S}_b \;\;: \;\; x \rightarrow x\;\;, \;\; t\rightarrow -t \;\; ,
\;\; {\rm \qquad \ if} \; \{E_s\;,\;\gamma = 0\} \;\; ;
\label{sym1}
\\
& \, & \hat{S}_c \;\; : \;\; x \rightarrow x+\frac{\lambda}{2} \;\;, \;\;
t \rightarrow -t \;\; , \;\;
{\rm \ if} \; \{f_{sh}\;,\;E_a\;,\;m=0\} \;\; .
\nonumber
\end{eqnarray}
%----------------------------------------------------------------------
Note that $\hat{S}_b$ and $\hat{S}_c$ require $\gamma=0$ (Hamiltonian case) 
and $m=0$ (overdamped case) respectively.
Another observation is that
all symmetries require certain symmetry properties of the external
drive $E(t)$, while the properties of the space-periodic force $f$
does not matter for $\hat{S}_b$. 
A proper choice of the drive may thus break all of the above 
listed symmetries with any coordinate dependence of the force $ \, f(x) $.

It is useful to provide an interpretation of the action of the
above symmetry operations on trajectories of (\ref{eqm1}). If the 
equation of motion is invariant under a symmetry, it implies that
a given solution or trajectory, when transformed using the symmetry
operation, yields again a solution or trajectory of the system.
It may be the same trajectory or a different one. The symmetry
$\hat{S}_a$ for $\gamma \neq 0$ transforms an attractor into
an attractor, repellers into repellers and basins of attraction
into basins of attraction. Thus if an attractor is mapped onto itself,
the average velocity on that attractor is zero. If an attractor is
mapped onto another attractor, we find complete symmetry between the
two attractors and their basins of attraction, while the average velocities
will be of opposite sign. 
The symmetry $\hat{S}_b$, which is valid for the Hamiltonian case,
simply relates two trajectories to each other, both having opposite
average velocities.

Once we add white noise, we have to consider some way of weighting
different trajectories. In that sense, if two different trajectories
are related by a symmetry $\hat{S}_a$ or $\hat{S}_b$ and have identical
statistical weight, their contributions to a total average current will
annihilate.

The symmetry $\hat{S}_c$ is acting in the overdamped limit, yet it is reverting
the sign of time. Thus it maps an attractor on a repeller and the
basin of attraction of an attractor into the basin of repulsion of a 
repeller. It follows that due to the low dimension $d=2$ of the
phase space of this problem, the compactness of the phase space and
the uniqueness of a trajectory running through a point in it, 
that all trajectories have zero average velocity when $\hat{S}_c$
is valid. 

In order to correctly incorporate the effects of noise, kinetic
equations have to be considered. This is done in Appendix A.

If at least one of the symmetries (\ref{sym1}) is valid, we conclude
that the averaged current is zero if an additional white Gaussian
noise term is added. This noise will simply lead to an exploration of 
the whole phase space and thus realize on finite times different trajectories 
of the deterministic system.

Once all of the above symmetries are broken, we may expect that in general
the average current will be nonzero. The understanding of the mechanisms
which will lead to this current can be best obtained in the absence
of noise for the deterministic system.

%######################################################################
\subsection{Rectification mechanisms in the Hamiltonian limit}
%######################################################################

As it was shown in \cite{kinetic} the value of the average current
is strongly enhanced for the underdamped case, when the
dissipation rate tends to zero. Thus we start our study of the
mechanisms with the dissipationless case $\gamma=0$. This Hamiltonian 
limit of Eq. (\ref{7}) is generically characterized 
by nonintegrable dynamics 
with a mixed phase space containing 
both chaotic and regular areas \cite{Zas}.
A stochastic layer appears around the separatrix of the 
integrable nondriven 
[$ \, E(t) \, = 0 $] Eq. (\ref{7}). 

Contributions to a nonzero current may come from trajectories
inside the stochastic layer as well as from regular unbound motion.
Because the latter always exist for motion in both directions,
strong effective cancellation of currents takes place \cite{symmetry},
while the strongest remaining contribution again comes from
the stochastic layer, which will be considered below.

\begin{figure}[htb]
\vspace{1.0pt}
\centerline{
\epsfig{figure=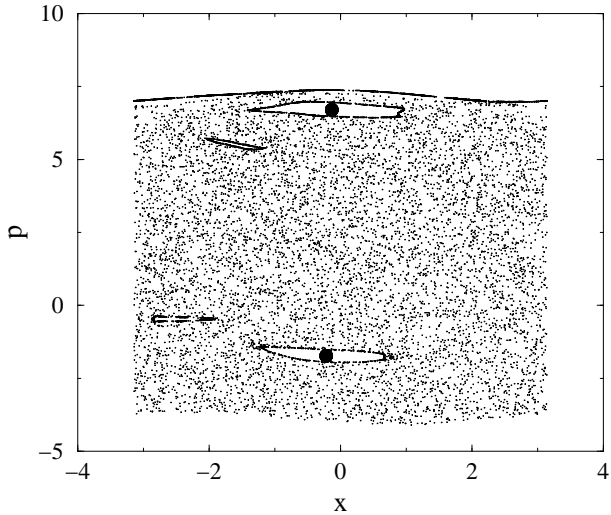,width=3.2in,angle=0}
}
\vspace{4.5pt}
 
\caption{Poincare map of the main stochastic layer of
(\ref{eqm1}) with $\gamma=0$ and functions (\ref{par1})
with $E_1=3.26$, $E_2=1.2$, $\alpha=\pi/2$ and $\omega=1$.  
The two filled circles correspond to the two observed limit cycle
attractors in the weakly dissipative case with damping
 $\gamma=10^{-4}$.}
\label{fig1}
\end{figure}

\begin{figure}[htb]
\vspace{1.0pt}
\centerline{
\epsfig{figure=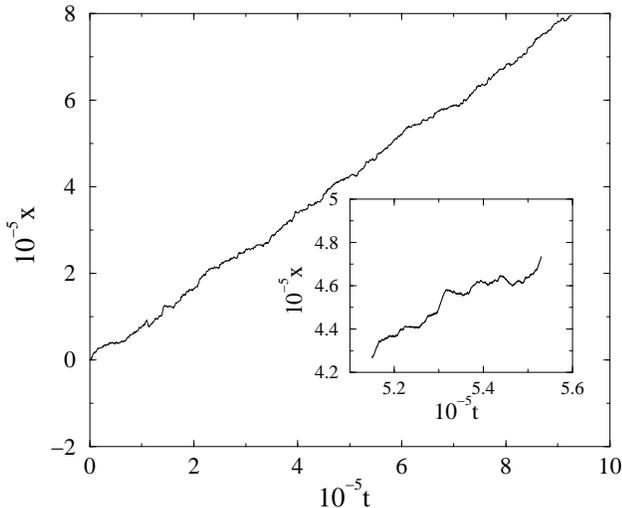,width=3.2in,angle=0}
}
\vspace{4.5pt}

\caption{$x(t)$ of trajectory from Fig.\ref{fig1}. 
Inset: Zoom of $x(t)$. Note the axis scales.}
\label{fig2}
\end{figure}

In the following we choose the following functions and parameters:
%------------------------par1------------------------------------------
\begin{eqnarray}
  V(x) & = & -\cos x \; , \label{par1} \\
  E(t) & = & E_1 \cos \omega t + E_2 \cos (2\omega t+\alpha) \; . 
\nonumber
\end{eqnarray}
%------------------------par1------------------------------------------
This choice ensures that the symmetries $\hat{S}_a$ and $\hat{S}_b$
are broken. In Fig.\ref{fig1} we show the Poincare map of the main stochastic
layer of our model.
%#########################FIG 2########################################
%######################################################################
Points with coordinates $p=\dot{x}$ and  $x \; mod \; 2\pi$ are drawn after
each period of the drive. In Fig. \ref{fig2} the time dependence of the
coordinate $x(t)$ is shown.
%#########################FIG 3########################################
%######################################################################
We observe a drift in accordance with the symmetry breaking
\cite{symmetry}. The average velocity is approximately $\langle
\dot{x} \rangle \approx 0.85$. 

\begin{figure}[htb]
\vspace{1.0pt}
\centerline{
\epsfig{figure=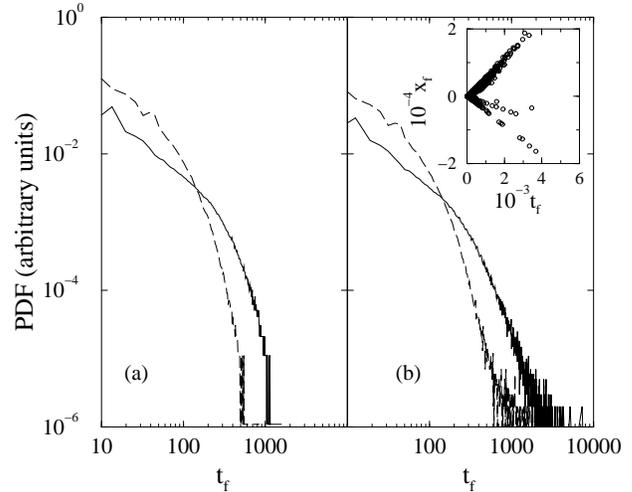,width=3.2in,angle=0}
}
\vspace{4.5pt}

\caption{
Probability distributions of ballistic flights to the right (solid line)
and left (dashed line). (a) - case with dissipation and noise.
(b) - Hamiltonian case.
Inset: Length of ballistic flight versus time of ballistic flight
for Hamiltonian case.
}
\label{fig3}
\end{figure}

In order to understand the
dynamical mechanisms of a nonzero dc current in the stochastic
layer, we first note that the stochastic layer is bounded in phase
space. The boundary contains of a fractal set of regular islands
embedded in the stochastic layer. A trajectory from the stochastic
layer may become trapped for quite long times in these boundaries
and perform ballistic-like (regular-like) dynamics.
The symmetry breaking of the equations of motion is reflected in a 
desymmetrization of these fractal boundary structures for the upper
and lower boundaries. This in turn leads to a desymmetrization of
distribution functions which characterize the probability to stick and
stay in such a boundary region \cite{last}.
In Fig.\ref{fig3}(b) the probability distribution functions
of the flight durations to the left and the right (for technical
details see \cite{last}) are shown. These functions are characterized
by algebaically decaying tails, and we clearly observe the above mentioned
desymmetrization.
%#########################FIG 4########################################
%@@@@@@@@@@@@@@@@@@@@@@@@@@@@@@@@@@@@@@@@@@@@@@@@@@@@@@@@@@@@@@@@@@@@@@
In the inset of 
Fig.\ref{fig3}(b) each point denotes the distance 
covered in a given flight during
the time of ballistic-like motion. We observe a fine structure
with three branches. Two major
ones correspond to the main ballistic channels
with opposite velocities while and a minor third one
corresponds to a channel with 
smaller negative velocity.

%######################################################################
%######################################################################
\subsection{Presence of dissipation}
%######################################################################
%#####################################################################

When dissipation is present in the system (i.e. $\gamma \neq 0$), the
phase space of the system separates into basins of attraction
of different low-dimensional attractors. Close to
the Hamiltonian limit these attractors are limit cycles:
$x(t+T)=x(t)+2 \pi m$, $\dot{x}(t+T)=\dot{x}(t)$, $m \in Z$. These limit cycles
are locked to the external periodic drive $E(t)$, therefore their
period $T$ is characterized by $T=2n\pi/\omega$, $n \in Z$. The
average velocity on the limit is given by
%---------------------------------------------------------------------
\begin{equation}
\langle \dot{x} \rangle =\frac{1}{T}\int_0^T {\dot x} dt=\frac{m}{n}\omega\;.
\label{12}
\end{equation}
%---------------------------------------------------------------------
When further away from the Hamiltonian limit chaotic attractors can 
appear via period-doubling bifurcations \cite{mateos}.

\begin{figure}[htb]
\vspace{1.0pt}
\centerline{
\epsfig{figure=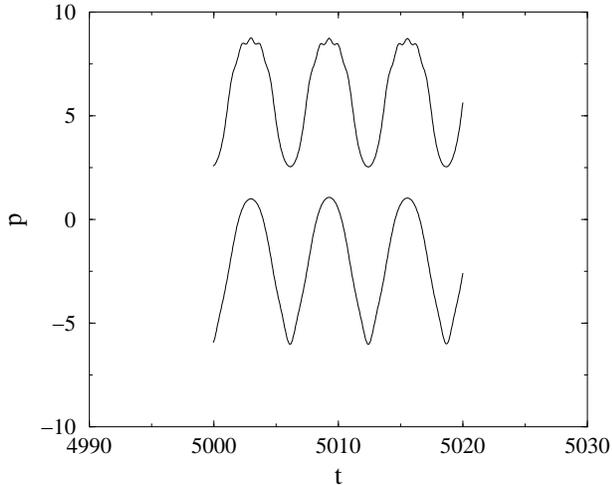,width=3.2in,angle=0}
}
\vspace{4.5pt}

\caption{$p(t)=\dot{x}(t)$ of both attractors for $\gamma=10^{-4}$
which correspond to the two filled circles in Fig. \ref{fig1}.}
\label{fig4}
\end{figure}

A numerical test reveals that in the case of  
$\gamma = 10^{-4}$ only two limit cycle attractors appear.
Their location in the Poincare map is shown in 
Fig. \ref{fig1}
%Kirpich_Dissipation_REsonance_1_Poincare.dat
%Kirpich_Dissipation_REsonance_2_Poincare.dat
by two filled circles. The dependence of $\dot{x}(t)$ on both
attractors is shown in Fig. \ref{fig4}.
%Kirpich_Dissipative_Low_Cycle_Trajectory.dat
%Kirpich_Dissipative_Upper_Cycle_Trajectory.dat
%Kirpich_Hamiltonian_Low_Elliptic_Point_Trajectory.dat
%Kirpich_Hamiltonian_Upper_Elliptic_Point_Trajectory.dat
%#########################FIG 5########################################
%######################################################################
Note that the attractors are 
located inside regular islands of the corresponding nondissipative system. 
These islands are characterized with nonzero winding numbers, 
and the sticking of 
the chaotic trajectory of the
nondissipative case provides with the above discussed ballistic
channels there. So we observe that the ballistic channels of the
Hamiltonian system 
survive by transforming into limit cycle attractors of the weakly
dissipative case.

\begin{figure}[htb]
\vspace{1.0pt}
\centerline{
\epsfig{figure=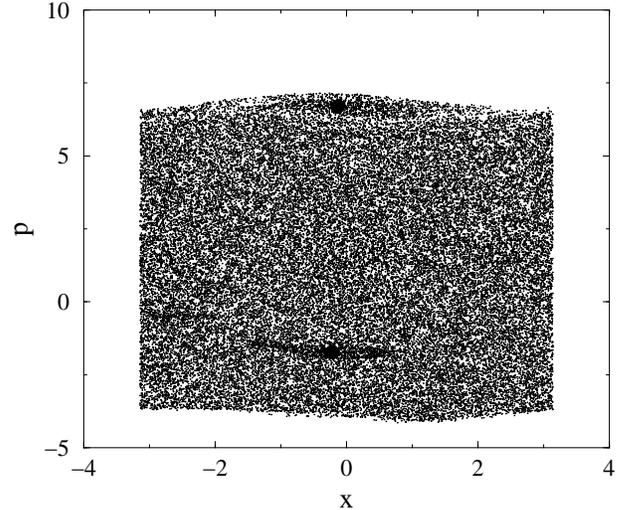,width=3.2in,angle=0}
}
\vspace{4.5pt}

\caption{Poincare map of the dissipative case with weak noise.
The two filled circles correspond to the two observed limit cycle
attractors for $\gamma=10^{-4}$.}
\label{fig5}
\end{figure}

As the attractors are located inside the stochastic layer of the Hamiltonian
limit, their basins of attraction are expected to show up with
a complex folding structure. This fact is manifested beautifully
if we add a noise of weak intensity
%P. E. Kloeden and E. Platen
%'Numerical Solution of Stochastic Differential Equations'
%(Springer NY 1992)
which corresponds to a small temperature $1/\beta=0.05$:
%----------------------------------------------------------------------
\begin{equation}
\langle \xi(t) \xi(t') \rangle = 2 \gamma /\beta \delta (t-t') \;\;.
\label{13}
\end{equation}
%----------------------------------------------------------------------
$ 1/\beta $ is small compared to the energy barrier of the periodic potential.
As it turns
out, in this case of strong external driving the system is so far from
the equilibrium case, that different scales have to be used in order
to compare with the noise intensity.
In Fig. \ref{fig5}
%Kirpich_Noise_Poincare.dat
we show the corresponding Poincare map over a total time
of $t_f = 10^6$. We observe that the trajectory is sticking for
long times to the two attractors. But most importantly we observe
frequent escapes from the attractors basins. Once the trajectory is
kicked out of such a basin, it starts to quickly explore the stochastic
layer space, due to the weak damping, weak noise, and the above mentioned
complex folding of the basin boundaries. This results in the fact
that the probability distribution of the velocities is far from being
Maxwellian.  
%#########################FIG 6########################################
%######################################################################

\begin{figure}[htb]
\vspace{1.0pt}
\centerline{
\epsfig{figure=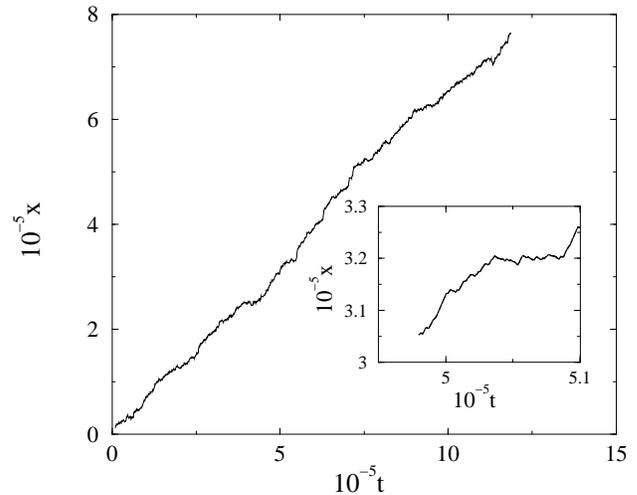,width=3.2in,angle=0}
}
\vspace{4.5pt}

\caption{$x(t)$ of trajectory from Fig.\ref{fig5} . Inset: Zoom 
of $x(t)$. Note the axis scales.}
\label{fig6}
\end{figure}

The dependence of $x(t)$ for the dissipative case with noise is shown
in Fig. \ref{fig6}.
%Kirpich_Noise_Trajectory.dat
%Kirpich_Noise_Trajectory_Zoom.dat
Although the dc current value has changed compared
to the Hamiltonian case, again the evolution is characterized by
sticking to ballistic channels.
%#########################FIG 7########################################
%######################################################################
An evaluation of the corresponding
ballistic flight time distributions is shown in Fig.\ref{fig3}(a).
A remarkable similarity to the case of the distributions for
the Hamiltonian case is observed. Also we observe that the power law
tails of the Hamiltonian case are replaced by exponential ones
in the dissipative case with noise, due to an expected
noise-induced cutoff in the
maximum correlation time.
%Histogram_Negative_Positive_Flights_Time.dat
%Histogram_Noise_Positive_Flights_Time.dat

A consequence of the above results is that
the loss of ballistic channels due to the variation of some
parameter may lead to a crossover-like decrease of the
current. In the next section we will design a model which does possess
these properties.

%#####################################################################
%#####################################################################
\section{Directed currents - the case of nonparabolic dispersion}
\label{NonParDR}
%#####################################################################
%#####################################################################

So far we have discussed the case of a classical particle with
parabolic dispersion, i.e. a kinetic energy quadratic in
the momentum $p$. What happens if we consider nonparabolic dispersion?
A prominent example would be a periodic dependence on $p$ as
for $(-\cos p)$, which reminds the consideration of a quantum
particle evolution in one band approximations.

%#####################################################################
\subsection{Absence of additional potential}
%#####################################################################

Let us first consider the case of a particle with a general
dispersion relation in the presence of additional ac driving:
%----------------------------------------------------------------------
\begin{equation}
H=\epsilon (p) - x E(t)
\;\;,\;\;\epsilon (p)=\epsilon (-p)\;\;.
\label{cosp}
\end{equation}
%----------------------------------------------------------------------
If the function $\epsilon(p)$ is chosen to be periodic, its
period is defined as $\lambda_p$.
Let us first consider the symmetries of the equations of motion
which change the sign of the velocity $\dot{x}$:
%----------------------------------------------------------------------
\begin{equation}
\dot{x} = \epsilon ' (p)\;\;,\;\; \dot{p} = E(t)\;\;.
\label{eqmcos}
\end{equation}
%----------------------------------------------------------------------
The following symmetries can be identified:
%----------------------------------------------------------------------
\begin{eqnarray}
&&x \rightarrow -x\; ,\; p \rightarrow -p\; ,\; t \rightarrow t+\frac{T}{2}
\;\; {\rm if} \;\; \{ E_{sh} \}\;,
\label{s1} \\
&&t \rightarrow -t\;,\; p \rightarrow -p\;\; {\rm if} \;\;\{ E_s \}\;,
\label{s2} \\
&&x\rightarrow -x \;,\; p \rightarrow p+\frac{\lambda_p}{2}
\;\; {\rm if} \;\; \{ \epsilon '_{sh} \}\;,
\label{s3} \\
&&t \rightarrow -t \; , \; p \rightarrow p+\frac{\lambda_p}{2}\;\;
{\rm if} \;\; \{ \epsilon '_{sh},E_a \}\;.
\label{s4}
\end{eqnarray}
%----------------------------------------------------------------------
Note that the last two operations (\ref{s3},\ref{s4}) may apply only
for periodic $\epsilon (p)$ functions. Furthermore these operations
change the energy of the undriven system. If these operations
connect different trajectories, they should not matter at finite temperatures
since different energies contribute with different weights.

Choosing the free particle case $\epsilon (p)=p^2/2$ we may arrive
at the conclusion that both relevant symmetries
(\ref{s1},\ref{s2}) can be violated by a proper choice of $E(t)$.
Yet, as shown in \cite{symmetry,yfr00pre}, the expected dc current should
be zero in such a case, in fact independent on the strength of
some additionally applied dissipation and noise. The case of
$\epsilon (p)=-\cos p$ is more involved. On one side it is
well known that the kinetic Boltzmann equation solution provides
with a nonzero dc current if both symmetries (\ref{s1},\ref{s2})
are broken \cite{kinetic}. On 
the other side this dc current tends to zero as the
dissipationless Hamiltonian case is approached. We will show in
the following that the reason for that is an additional symmetry
of the {\it solutions} of (\ref{eqmcos}) due to the integrability
of (\ref{cosp}).

Indeed, integration of (\ref{eqmcos}) yields
%----------------------------------------------------------------------
\begin{equation}
p(t)=g(t;p_0,t_0)=p_0+E_I(t)-E_I(t_0)\;.
\label{intp}
\end{equation}
%----------------------------------------------------------------------
where $\dot{E}_I(t) = E(t)$. The function $g$ has the property
%----------------------------------------------------------------------
\begin{equation}
g(x;y,z) = -g(z;-y,x)\;.
\label{symint}
\end{equation}
%----------------------------------------------------------------------
This symmetry operation relates points on a given trajectory to
points on a {\it set} of other trajectories. Such a symmetry
can not be derived from the equations of motion. It rather is a result
of the integrability and reflects the symmetry of the {\it solutions},
i.e. of the phase space flow (in contrast symmetries of equations of motion
relate two trajectories to each other).

The symmetry (\ref{symint}) is in place independent
of the choice of the functions $\epsilon(p)$ and $E(t)$
and provides with the following consequences.
If we consider a corresponding kinetic equation with finite
dissipation, the loss of correlations implies an averaging over
the initial phase $t_0$ of the field $E(t)$. This averaging persists
in the dissipationless limit, while on the other side we recover
the above considered Hamiltonian properties. The averaging over
$t_0$ leads to an enforcement of the symmetry (\ref{symint}),
which in turn by changing the sign of $p$ changes the sign
of $\dot{x}$. Thus a vanishing dc current is expected in that limit,
in accordance with numerical and analytical analysis of kinetic
equations. This result can be also obtained from (\ref{intp})
and (\ref{eqmcos}) if an averaging with a distribution function,
which is symmetric in $p_0$ and independent of $x$ and $t_0$,
is performed:
%----------------------------------------------------------------------
\begin{equation}
\int _0^T {\rm d} t \int_0^T {\rm d}t_0 \dot{x} |_{p_0} +
\int _0^T {\rm d} t \int_0^T {\rm d}t_0 \dot{x} |_{-p_0} = 0 \, .
\label{zero}
\end{equation}
%----------------------------------------------------------------------

To detect a dc current carried by the electrons in a single band one has to
break the integrability of the semiclassical equations of motion. We consider
this case in the next section. An alternative way is a much more careful
treatment of the quantum mechanical problem (see, for example \cite{meso})
which is however far beyond the scope of the present paper.

%######################################################################
\subsection{Presence of additional potential}
%######################################################################

Things change drastically if we add an external space-periodic
potential
to (\ref{cosp}).
Such a system is in general nonintegrable, and we may expect
the additional symmetry (\ref{symint}) to not be present.
To be precise, we choose the following model here:
%----------------------------------------------------------------------
\begin{equation}
H = -\cos p -\frac{1}{2}\cos (2p) - \frac{\xi^2}{3}
\cos \left (\frac{x}{\xi}\right ) - \frac{\xi}{3}x E(t)\;.
\label{scaledH}
\end{equation}
%----------------------------------------------------------------------
The corresponding equations of motion read
%----------------------------------------------------------------------
\begin{eqnarray}
\dot{x}&=& \sin p + \sin (2p)
\label{dotx}\;, \\
\dot{p} &=& -\frac{\xi}{3} \sin \left (\frac{x}{\xi} \right ) + 
\frac{\xi}{3}E(t)\;
\label{dotp}
\end{eqnarray}
%----------------------------------------------------------------------
The free parameter $\xi$ can be used to reach the limit of (\ref{eqm1}).
Indeed for $\xi \rightarrow 0 $ $\dot{p} \rightarrow 0$.
Choosing a small value of $|p| \ll 1$ will keep the momentum small.
This allows for an expansion of the right hand side of
(\ref{dotx}) to first order. Additional rescaling
$x \rightarrow x/ \xi$ and $p \rightarrow (3/\xi) p$ transforms
the problem exactly to Eq. (\ref{eqm1}) (with $\gamma=0$).
On the other side, a choice of a finite value of $p$ may
lead to trajectories which are not contained in Eq. (\ref{eqm1}).
In addition (\ref{scaledH}) is periodic in $p$, so the phase
space is compact.

Let us discuss the relevant symmetries of Eqs. (\ref{dotx})-(\ref{dotp}).
Due to the presence of the $\cos (2p)$ term in (\ref{scaledH})
the shift symmetry of the kinetic energy is broken
(in terms of electrons this implies loosing of the particle-hole
symmetry). We can identify
the following symmetry operations which lead to a change of sign
of $\dot{x}$:
%--------------------------sc1-sc2-------------------------------------
\begin{eqnarray}
&& x\rightarrow -x\;,\; p \rightarrow -p \;,\;
t \rightarrow t+\frac{T}{2}\;\;{\rm if} \;\;
\{ E_{sh} \};\,
\label{sc1} \\
&& t \rightarrow -t\;,\; p \rightarrow -p \;\;
{\rm if} \;\;
\{ E_s \}\;.
\label{sc2}
\end{eqnarray}
%----------------------------------------------------------------------
In the following we choose $E(t)$ from (\ref{par1}) which ensures
that both symmetries are broken.

 For $\xi \ll 1$ the phase space of (\ref{scaledH})
is characterized by the presence of two disconnected stochastic layers.
One of them corresponds to the layer in Fig. \ref{fig1}.
%#########################FIG 9########################################
\begin{figure}[htb]
\vspace{1.0pt}
\centerline{
\epsfig{figure=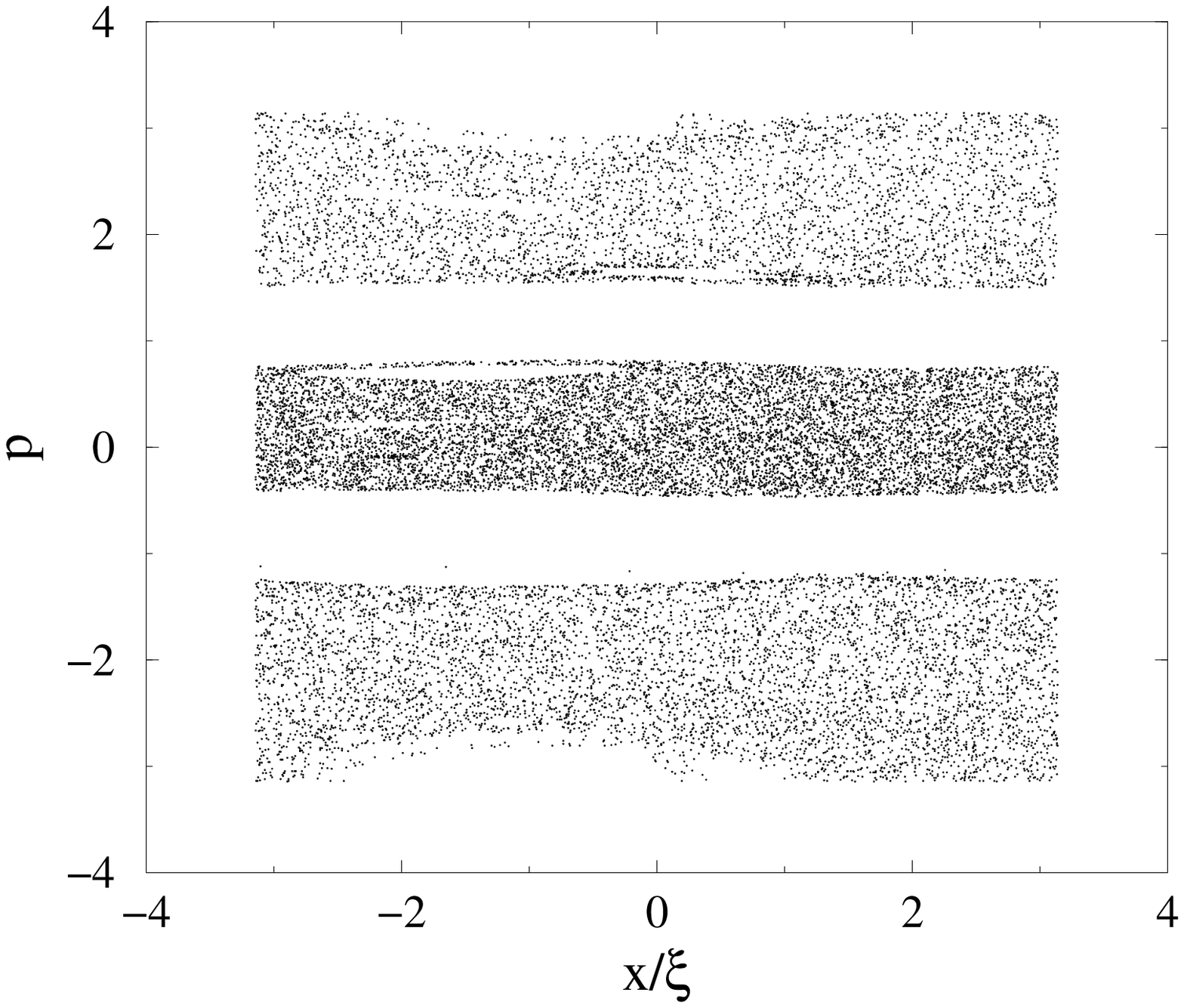,width=3.2in,angle=0}
}
\vspace{4.5pt}

\caption{Poincare map of Eqs. (\ref{dotx})-(\ref{dotp}) for $\xi=0.35$.}
\label{fig7}
\end{figure}
%######################################################################
In Fig. \ref{fig7} we show the Poincare map of (\ref{scaledH}) for
$\xi=0.35$. Note that the central layer is the one which
continuously transforms into the layer in Fig. \ref{fig1} when $\xi
\rightarrow 0$. An increasing of $\xi$ will lead to a merging of
the layers, which is followed by losses (closings) of
ballistic channels, that are located near the boundaries of the
central  stochastic layer. The Poincare map for $\xi=1$ is shown
in Fig. \ref{fig8}.
%#########################FIG 10#######################################
\begin{figure}[htb]
\vspace{1.0pt}
\centerline{
\epsfig{figure=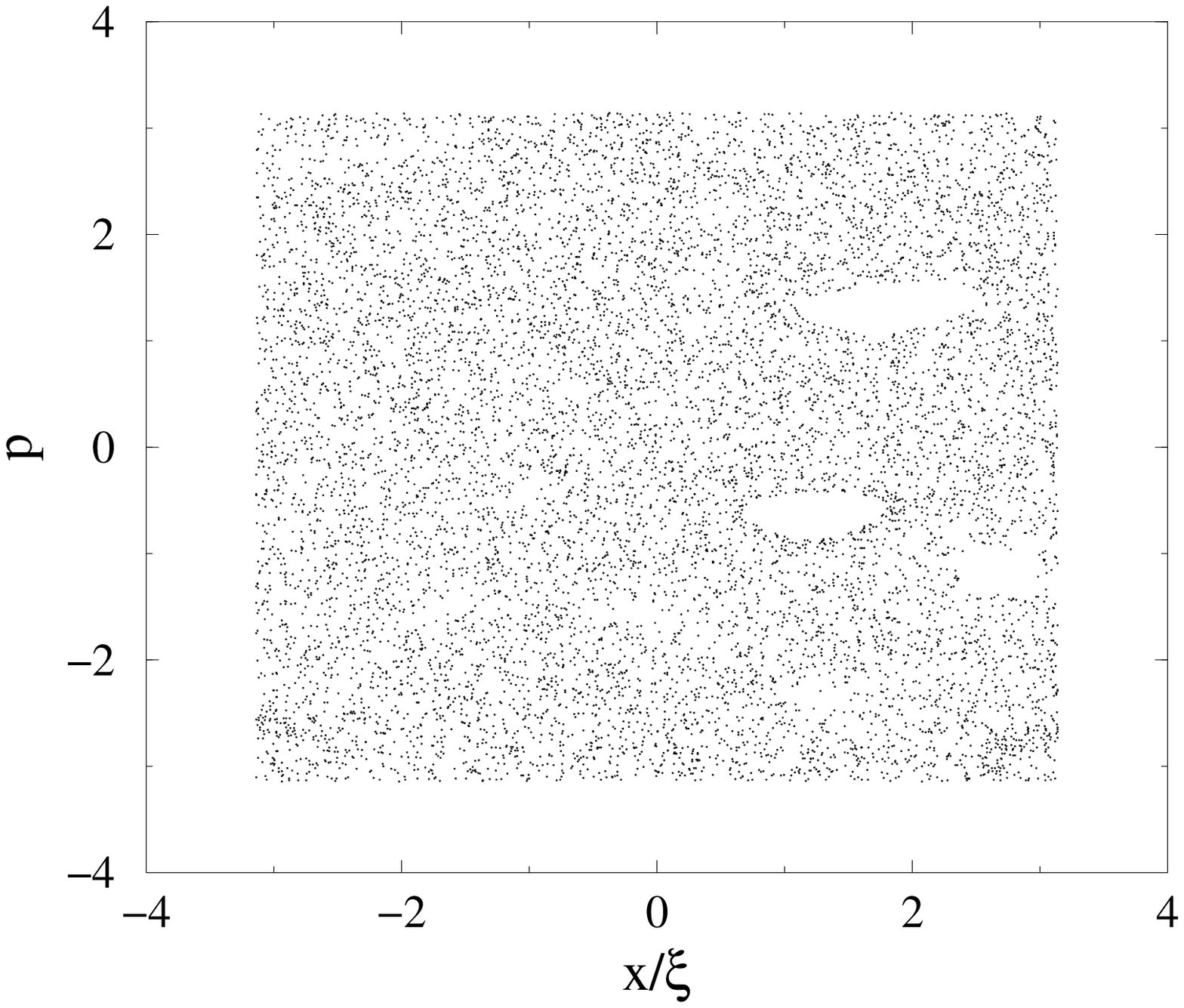,width=3.2in,angle=0}
}
\vspace{4.5pt}

\caption{Poincare map of Eqs. (\ref{dotx})-(\ref{dotp}) for $\xi=1$.}
\label{fig8}
\end{figure}
%######################################################################
Let us discuss a trajectory with initial conditions $p=0$ and
$x=\xi \pi$, which ensures that we always start in the stochastic
layer which corresponds to Fig.1 in the limit of small $\xi$.
The dependence of the
scaled dc current $j_s = \langle \dot{x} \rangle / \xi$ is 
shown in Fig. \ref{fig9} as
a function of $\xi$.
Specifically its values for the presented Poincare maps are
$j_s \approx 0.63$ for $\xi=0.35$ and $j_s \approx 0.002 $ for $\xi=1$.
We observe indeed that
the current variations correspond to a closing or opening of ballistic
channels.
%#########################FIG 11########################################
\begin{figure}[htb]
\vspace{1.0pt}
\centerline{
\epsfig{figure=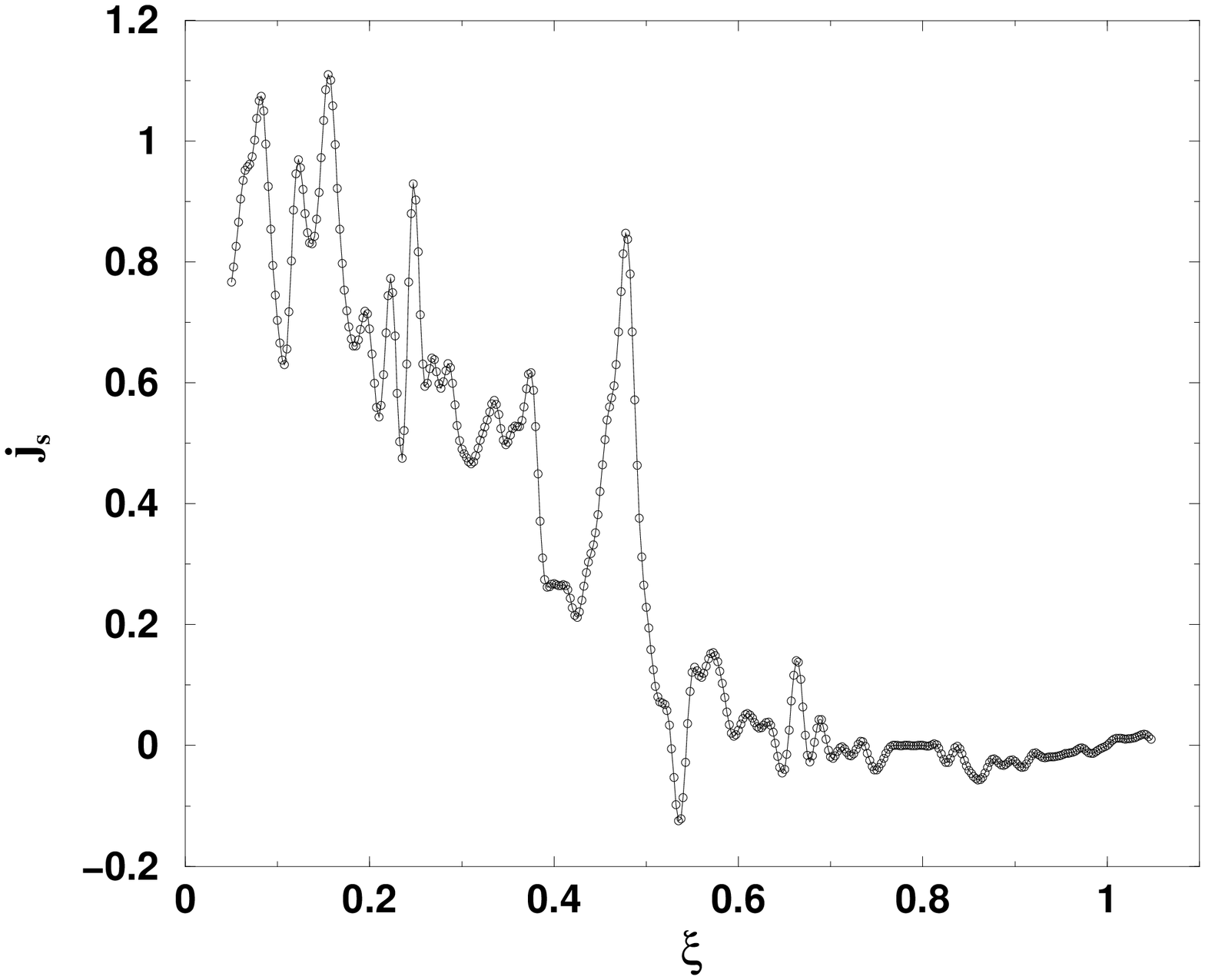,width=6.75cm,angle=270}
}
\vspace{4.5pt}

\caption{$j_s$ as a function of $\xi$. Note that the error in determining
the current value is of the order of 10\%. A fine structure of pronounced
peaks is connected with the opening of new ballistic channels 
due to the overlap 
of the stochastic layer with high-order resonances.
}
\label{fig9}
\end{figure}
The above results support the general expectation that in the Hamiltonian
limit a mixed phase space is needed, so that the current vanishes both
in the case of an integrable system as well as in the case of a fully
chaotic one.

%######################################################################
%######################################################################
%######################################################################
\section{Conclusions}
%######################################################################
%######################################################################
\label{Conclusions}

We presented a symmetry approach to the effect of rectification
due to external ac fields applied to a low-dimensional dynamical
system with optional contact to a heat bath. 
The nonlinear (nonadiabatic) response is used to explain the
effect. 
We explained the
mechanisms of such a rectification for different cases. 
In the case of a particle moving in a space-periodic potential
the explanation is given in terms of desymmetrization of 
ballistic channels which correspond to motion in different 
directions.
While this explanation starts from the case of a nondissipative
Hamiltonian limit, we showed that the ballistic channels
are robust with respect to adding dissipation and fluctuations.
The mechanism of the directed current in the presence of dissipation
is hidden in the desymmetrization of the basins of attraction
of previously symmetry-related limit cycles with oppositely
directed velocities.

A recent geometric approach by H. Schanz et al in \cite{columbia}
provides with an elegant way to account for the average drift
velocity in the Hamiltonian case. The basic ingredient of this
approach is the assumption of an invariant density distribution in
a stochastic layer being constant inside the layer. The resulting
sum rule, when evaluated for concrete cases, provides with average
velocities which quantitatively agree with our observations. A
consequence of the sum rule is that a nonzero directed current may
appear only in systems with a mixed phase space. This result
correlates with our discussion of the mechanisms of directed
currents in terms of ballistic channels, which may appear only
provided the stochastic layer has some bounds with regular phase
space regions, and the ballistic channels occur precisely in the
neighborhood of these bounds. Open questions are: i) is there
purely diffusive (chaotic) directed motion besides of transport
through ballistic 
channels? (ii) when can one of these two mechanisms prevail?

It was shown that
the symmetry properties of the Fokker-Planck equation
match the symmetries of the corresponding
deterministic equations of motion in the absence of noise. 
Similar results hold for a Boltzmann equation \cite{kinetic}.

Our symmetry considerations may be as well used to explain
rectifications in such diverse situations as i) the directed
motion of particles in non-Newtonian liquids \cite{vs85pla},
 ii) the appearance of ring currents
for particles moving in a two-dimensional space-periodic potential
\cite{indus,d-c-n98pf}, iii) the appearance of directed  heat flux 
currents in systems of interacting particles \cite{fzmf01}, iv) the 
appearance of a nonzero dc polarization,
v) quantum ratchets \cite{vinokur}, to name a few. 
Symmetry analysis is also instructive in the case of
less conventional models, like systems with built-in
asymmetry \cite{d02pla,Hu,Floria}, where the asymmetry 
is hidden in a many body
system's internal interactions.
\\
\\
Acknowledgments
\\
We thank T. Dittrich, 
M. V. Fistul, P. H\"anggi, R. Ketzmerick, A. Miroshnichenko, P. Reimann
and H. Schantz for stimulating discussions, and A. Miroshnichenko for
numerical help with solving Fokker-Planck equations.
\\
\\
\\
\appendix

{\bf Appendix A}
\\

The Fokker-Planck equation for the probability distribution $W(x,p,t)$
with $p\equiv \dot{x}$ of equations (\ref{7}),(\ref{7-2}) reads
\cite{Reimann}
\begin{equation}
\frac{\partial W}{\partial t} =
-p \frac{\partial W}{\partial x} + \frac{1}{m}\frac{\partial}{\partial p}
\left[ -f(x)-E(t) +\gamma p + \frac{\gamma}{m \beta}  
\frac{\partial}{\partial p} \right]
W\;.
\label{FP}
\end{equation}
This equation is linear in $W$,
preserves the norm $\int W dx\; dp$ and is dissipative. For a fixed norm
any initial condition will converge to a single attractor solution $W_s$.
For the case $E(t)=0$ it is easy to see that the attractor is the
Gibbs distribution. For nonzero $E(t)$ the attractor solution $W_s$
will be periodic in $x$ and $t$. The average current is given by
\begin{equation}
\langle p \rangle  = \int p W_s dx\;dp\; dt\;.
\label{tok1}
\end{equation}
It follows that (\ref{FP}) is invariant under the following transformations:
\begin{eqnarray}
& \, & \hat{S}_a\;: \; x \rightarrow -x\;, \; p \rightarrow -p\;,\; 
t\rightarrow t+\frac{T}{2}\;,
\; {\rm if} \; \{f_{a}\;,\; E_{sh} \} \; ;
\nonumber
\\
& \, & \hat{S}_b \;: \; x \rightarrow x\;, \; t\rightarrow -t \; ,
\; p \rightarrow -p\;,\; {\rm \qquad \ if} \; \{E_s\;,\;\gamma = 0\} \; .
\label{sym1-FP}
\end{eqnarray}
At the same time (\ref{tok1}) is changing sign. Since the solution
$W_s$ is unique, the conclusion is that $ \langle p \rangle =0$.

The overdamped limit $m=0$ is described by the following
Fokker-Planck equation \cite{Reimann}
\begin{equation}
\gamma \frac{\partial W}{\partial t} =
\frac{\partial}{\partial x} \left[ -f(x)-E(t) \right]W + \frac{1}{\beta}
\frac{\partial ^2 W}{\partial x^2}\;.
\label{FPOD}
\end{equation}  
The average current is given by
\begin{equation}
\langle p \rangle = \int f(x) W_s dx\;dt\;.
\label{tok2}
\end{equation}
The symmetry $\hat{S}_a$ is again holding. However the symmetry
\begin{eqnarray}
& \, & \hat{S}_c \; : \; x \rightarrow x+\frac{\lambda}{2} \;\;, \;\;
t \rightarrow -t \;\; , \;\;
{\rm \ if} \; \{f_{sh}\;,\;E_a\;,\;m=0\} \;\; .
\nonumber
\end{eqnarray}
does not follow from (\ref{FPOD}) in a similar way. 
It is interesting to note that Reimann \cite{Reimann01}
considered the symmetries of the original stochastic differential
equation (\ref{7}),(\ref{7-2}) and argued for a holding of 
$\hat{S}_c$ for nonzero noise intensities. 
Fistul \cite{mf02-2} showed that $\hat{S}_c$ connects the original
Fokker-Planck equation (\ref{FPOD}) with its conjugated counterpart.

Here we will use a related approach and prove the vanishing of (\ref{tok2})
when $\hat{S}_c$ holds. First we remind that the distribution $W_s$
is periodic in $x$ and $t$. Next we note that the operators $\partial/\partial
t$
and $\partial / \partial x$ are anti-Hermitian on the space of $x,t$-periodic
functions. Let us define the operator $T$ as
\begin{equation}
T = \frac{-\frac{\partial}{ \partial x}}{\gamma \frac{\partial}{\partial t}
-\frac{\partial ^2}{\beta \partial x^2} + E(t) \frac{\partial}{\partial x}}
\;.
\end{equation}
Then (\ref{FPOD}) can be rewritten as a Lippmann-Schwinger type integral
equation:
\begin{equation}
W = 1 + T f(x) W \;.
\label{ls}
\end{equation}
Provided the conditions for $\hat{S}_c$ hold, i.e.
$E(t) \equiv E_a(t)$, the operator
$T$ has the property 
\begin{equation}
T^{\dagger} = - T(-t)\;,\;T(x+x_0) = T(x)\;.
\label{propertyT}
\end{equation}
Expanding (\ref{ls}) in a formal series in $f(x)$ we obtain
\begin{equation}
W_s = 1 + \sum_{n=1}^{\infty} (T f(x))^n\;.
\end{equation}
With (\ref{tok2}) we obtain for the average current
\begin{equation}
\langle p \rangle = \sum_{n=1}^{\infty} \int f(x) (T f(x))^n dx\;dt\;.
\label{series}
\end{equation}
Since all terms in (\ref{series}) are real valued, we conclude
that all integrals with even $n$ vanish because of $f(x) \equiv f_{sh}(x)$
and these terms contain odd powers of $f$.
All integrals with odd $n$ vanish because of 
(\ref{propertyT}), since they contain odd powers of $T$. 
Thus we find that indeed the average current exactly
vanishes when $\hat{S}_c$ holds.

We also 
performed a numerical solution of (\ref{FPOD}) for $T=\gamma=1$
and $f(x)=\sin x$, $E(t)=\sin (\omega t) + \sin (2\omega t+\alpha)$
with $\omega=0.8$.
The function $W$ was expanded in a Fourier series in $x$ and twenty
Fourier harmonics have been taken into account. The resulting coupled
ordinary differential equations were simply integrated until the system
reached the final attractor $W_s$.
\begin{figure}[htb]
\vspace{1.0pt}
\centerline{
\epsfig{figure=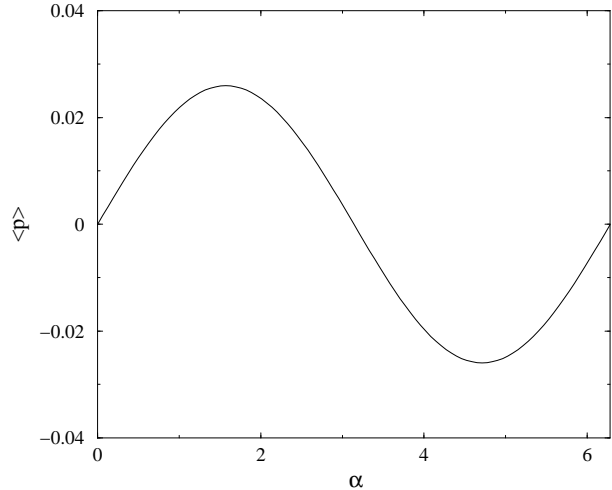,width=6.75cm,angle=270}
}
\vspace{4.5pt}

\caption{$\langle p \rangle$ as a function of $\alpha$ (see text). 
}
\label{fig10}
\end{figure}
For the above choice of functions $f(x)$ and $E(t)$ the symmetry $\hat{S}_a$
is violated because $E(t)$ is not shift-symmetric. At the same time
the symmetry $\hat{S}_c$ is violated for all values of $\alpha$ except
for $\alpha=0,\pm \pi, \pm 2\pi,...$. In Fig. \ref{fig10} the dependence
of the average current $\langle p \rangle$ (\ref{tok2}) on $\alpha$ is
shown. We observe that for $\alpha=0,\pi, 2\pi$ the symmetry $\hat{S}_c$
is restored and the current is vanishing (the 
absolute numerical value is less than
$10^{-12}$).


\begin{thebibliography}{99}

\bibitem{Reimann}
P. Reimann, Phys. Rep. {\bf 361}, 57 (2002).

\bibitem{symmetry}
S. ~Flach, O.~Yevtushenko, and Y.~Zolotaryuk, Phys. Rev. Lett. {\bf
84}, 2358 (2000).

\bibitem{kinetic}
O. Yevtushenko, S. Flach, Y. Zolotaryuk and A. A. Ovchinnikov,
Europhys. Lett. {\bf 54}, 141 (2001).

\bibitem{fzmf01} S. Flach, Y. Zolotaryuk, A. Miroshnichenko and 
M. V. Fistul, {\bf nlin.PS/0110013}, to appear in Phys. Rev. Lett.

\bibitem{fo01pa}
S. Flach, A. A.  Ovchinnikov, Physica A {\bf 292}, 268 (2001);
S. Flach, A. Miroshnichenko and A. A. Ovchinnikov, Phys. Rev. B {\bf 65},
104438 (2002).

\bibitem{aao02}
A. A. Ovchinnikov, {\bf cond-mat/0110616}.

\bibitem{multi-valued}
Underdamped systems often display hysteretic dependence on
external fields, which leads to multi-valued response functions.
In that case the overall symmetry of a response function is not
sufficient to conclude about the current rectification. The case
of (\ref{eqm1}),(\ref{par1}) in the adiabatic limit is a nice example.
For $\alpha =\pi/2$ the ac drive is antisymmetric, yet a nonzero
rectification will take place, in full accord with the symmetry
analysis of the underlying equations of motion.

\bibitem{bh94}
R. Bartussek, P. H\"anggi and J. G. Kissner,
Europhys. Lett. {\bf 28}, 459 (1994).

\bibitem{maf02}
M. V. Fistul, {\bf cond-mat/0010087}, Phys. Rev. E, in print.

\bibitem{vinokur} S. Scheidl and V. M. Vinokur,
{\bf cond-mat/0201008}, (2002).

\bibitem{Zas}
G.~M. Zaslavsky,
\newblock {\em Physics of chaos in Hamiltonian systems}.
\newblock (Imperial College Press,  1998).

\bibitem{last}
S.~Denisov and S.~Flach, Phys. Rev. E {\bf 64}, 056236 (2001);
S.~Denisov, J. Klafter, M. Urbakh and S.~Flach, submitted
to \newblock{Physica D}, (2001) .

\bibitem{mateos}
J. ~Mateos, Phys. Rev. Lett. {\bf 84}, 258 (2000).

%\bibitem{Ott}
%E. Ott, {\em Chaos in Dynamical Systems} (Cambridge University
%Press, Cambridge, 1993)

\bibitem{yfr00pre}
O. Yevtushenko, S. Flach and K. Richter, Phys. Rev. E {\bf 61}, 7215
(2000). 

\bibitem{meso}
M. Vavilov, V. Ambegaokar, I. Aleiner, Phys. Rev. B {\bf 63}, 
195313 (2001).

%\bibitem{comment1} Note that the problem of collective description
%of the rectification can be done, alternatively, using the
%Liouville equation for the probability distribution function
%in the corresponding three-dimensional phase space. The
%equations in both cases would look the same.

%\bibitem{resibois}
%P.~R\'esibois and M. De Leener,
%\newblock {\em Classical Kinetic Theory of Fluids}.
%\newblock (A Wiley-Interscience Publication,  1977).

\bibitem{columbia}
H. Schanz, M.-F. Otto, R. Ketzmerick and T. Dittrich, 
Phys. Rev. Lett. {\bf 87} 
(2001), 070601.


\bibitem{vs85pla} A. K. Vidybida and A. A. Serikov, 
Phys. Lett. A {\bf 108}, 170 (1985).

\bibitem{indus}
A. W. Ghosh and S. V. Khare, Phys. Rev. Lett, {\bf 84}, 5243
(2000).

\bibitem{d-c-n98pf} D. del-Castillo-Negrete, 
\newblock{Physics of Fluids} {\bf 10}, 576 (1998).

\bibitem{d02pla} S. Denisov, {\bf nlin.CD/0007034}, to
appear in \newblock{Phys. Lett. A}, (2002).

\bibitem{Hu}
Z. Zheng, G. Hu, B. Hu, Phys. Rev. Lett. {\bf 86}, 2273 (2001).

\bibitem{Floria}
S. Cilla, F. Falo and L. M. Floria, Phys. Rev. E {\bf 63}, 031110
(2001).

\bibitem{Reimann01}
P. Reimann, Phys. Rev. Lett. {\bf 86}, 4992 (2001).

\bibitem{mf02-2}
M. V. Fistul, private communication.

\end{thebibliography}
\end{document}